\documentstyle[12pt]{article}
\def\cmm2{{\,\rm cm^{-2}}}
\def\cm2{{\,{\rm cm}^2}}
\def\cmm3{{\,{\rm cm}^{-3}}}
\def\gcmm3{{\,{\rm g\,cm^{-3}}}}

\newbox\bigstrutbox
\setbox\bigstrutbox=\hbox{\vrule height12pt depth5pt width0pt}
\def\bigstrut{\relax\ifmmode\copy\bigstrutbox\else\unhcopy\bigstrutbox\fi}

\def\la{\mathrel{\mathpalette\fun <}}
\def\ga{\mathrel{\mathpalette\fun >}}
\def\fun#1#2{\lower3.6pt\vbox{\baselineskip0pt\lineskip.9pt
  \ialign{$\mathsurround=0pt#1\hfil##\hfil$\crcr#2\crcr\sim\crcr}}}

\font\teni=cmmi10 scaled 1200
\font\seveni=cmmi7 scaled 1200
\font\fivei=cmmi5 scaled 1200
\font\tensy=cmsy10 scaled 1200
\font\sevensy=cmsy7 scaled 1200
\font\fivesy=cmsy5 scaled 1200

\skewchar\teni='177 \skewchar\seveni='177 \skewchar\fivei='177
\skewchar\tensy='60 \skewchar\sevensy='60 \skewchar\fivesy='60

\font\ninei=cmmi10
\font\sixi=cmmi7
\font\fouri=cmmi5
\font\ninesy=cmsy10
\font\sixsy=cmsy7
\font\foursy=cmsy5

\skewchar\ninei='177 \skewchar\sixi='177 \skewchar\fouri='177
\skewchar\ninesy='60 \skewchar\sixsy='60 \skewchar\foursy='60

\outer\def\beginsection#1\par{\bigskip\vbox{\message{#1}\noindent{\bf#1}}
  \nobreak\smallskip\vskip-\parskip\noindent}
\def\exdent#1\par{\noindent\hang\frenchspacing#1\par}
\begin{document}
\begin{flushright}
astro-ph/9408061
\end{flushright}
\vskip 0.5in
\begin{center}
{\Large \bf A Merger Model and Globular Cluster Formation}\\
\vskip 1 cm
{\bf Sangjin Lee$^1$, David N.~Schramm$^{1,2}$, \& Grant J.~Mathews$^3$}\\
\smallskip
{\em $^1$Department of Physics, The University of Chicago}\\
{\em Chicago, IL 60637}\\
\smallskip
{\em $^2$NASA/Fermilab Astrophysics Center,}\\
{\em Fermi National Accelerator Laboratory}\\
{\em Batavia, IL 60510}\\
\smallskip
{\em $^3$University of California, Lawrence Livermore National Laboratory}\\
{\em Livermore, CA 94550}\\
\vskip 0.5in
{\bf Abstract}
\end{center}
\bigskip
We propose a self-consistent model for globular cluster formation in, but not
limited to, our Galaxy, based on the merger model of Mathews \& Schramm
(1993). Stars and star clusters form in bursts at the merging interfaces as
protogalactic clouds collide. We describe the
formation of those star clusters with a simple schematic ansatz which takes
into account the thermal and Kelvin-Helmholtz instabilities. It is shown
that this model is consistent with many observational properties such as the
age and metallicity distributions of globular clusters, the overall
number of globular clusters, and the near constancy of the number of globular
clusters in different size host galaxies. Most of the features of this merger
model are insensitive to choices of parameters. However, the model does not
exhibit two distinct populations of globular clusters, i.e. halo clusters and
disk clusters. Possible explanations for this are presented.\\
\vskip 0.5in
\noindent{\em Subject headings}: cosmology: theory - galaxies: star clusters -
Galaxy: formation - Galaxy: globular clusters: general - stars: formation

\bigskip

\noindent{$^{\ast}$} Submitted to {\em The Astrophysical Journal}\\

\newpage
\section{Introduction}
Globular clusters have the focus of considerable study
both in cosmology and in astrophysics. Since they are
inferred to be among the oldest objects in our Galaxy, their age serves as a
lower bound on the age of the Galaxy and the Universe (Chaboyer et al. 1992a;
Chaboyer, Sarajedini, \& Demarque 1992b; Sandage 1993). The dating of
globular clusters has yielded interesting bounds on cosmological parameters,
such as the interrelationship between present fraction of the closure density
of the Universe $\Omega_0$, and the Hubble constant $H_0$.
Since at least some of globular clusters are extremely old, they also provide
information about the early galactic environment and galaxy formation process.

There have been several different models for the origin of globular clusters.
Peebles \&
Dicke (1968) first pointed out that globular clusters might
have formed even before the collapse of the protogalaxy, noting the fact that
the baryonic Jeans mass right after decoupling is about the size of a globular
cluster. However, it cannot explain why there are so few intergalactic globular
clusters, if any, and why the properties of globular clusters are correlated
with host galaxies. Fall \& Rees (1985)
suggested that globular clusters may have formed out of thermal instabilities
during the collapse of the protogalaxy. In their model, cold dense clouds
condense out of hot and tenuous background to form as progenitors of globular
clusters. While many theories on globular
cluster formation assume a smooth and rapid
collapse of the protogalaxy (see e.g. Eggen, Lynden-Bell, \& Sandage 1962),
there
are indications (Searle \& Zinn 1978) that the early Galactic environment might
have been much more chaotic and violent.

Murray \& Lin (1992) have argued that
self-gravitating
clouds are unstable to fragmentations and spontaneous star
formation so that globular clusters must form from sub-Jeans-mass clouds.
In subsequent work Murray et al. (1993) have shown that only clouds in a
limited mass range ($10^4 M_{\odot} \la M \la 10^6 M_{\odot}$) can survive
both the
Kelvin-Helmholtz (KH) instability (which disrupts the clouds as they move
through the hot background medium) and the thermal instability (i.e.
spontaneous cooling and star formation)
which does not lead to bound clusters. Clouds within the critical mass range
will form globular clusters if they are induced into cooling and collapse by
collisions of sufficient velocity. In what follows we develop a simple
schematic model of protogalactic mergers, to describe globular cluster
formation and metallicity within this context.

Mathews \& Schramm (1993, called MS hereafter) proposed a schematic merger
model
for the formation of the halo and chemical evolution of the Galaxy in which
the protogalaxy forms by
mergers of small subgalactic gas clouds.
The mergers during the collapse of the protogalaxy can produce a substantial
number of stars in addition to the normal star formation activity within the
gas
clouds. Using this merger model, they could provide insight into
a number of problems, such as the apparent discrepancy of
the various age
estimators of the Galaxy, the G-dwarf problem, etc.

The formation and evolution of globular clusters arises quite naturally in the
framework of such a merger model. Here we explore the consequences of
this merger model on the formation of globular clusters.

\section{Formalism}

An exact analytic treatment of the merging processes is impossible. A detailed
physical description requires
numerical hydrodynamic simulations using widely varying initial conditions,
etc. (e.g. Lin \& Murray 1992; Brown, Burkert, \& Truran 1991)
The next best, however, is suppose a reasonable schematic model which
encompasses the basic features of the large scale numerical simulations.

Following the simple kinematical argument given in MS, the merger rate per
protogalactic
cloud can be approximated
\begin{equation}
\lambda_m = \frac{1}{2} \frac{N-1}{V} \sigma v
\left(1-e^{-\frac{t-t_0}{t_{\em vir}}}\right)
\end{equation}
where $N$ is the number of clouds, $V$ is the volume of the protogalaxy,
$\sigma$ is the merger cross section, which is approximated by a geometric
cross section, and $v$ is the virial velocity which is related to the median
radius of the protogalaxy by
\begin{equation}
v^2 \simeq 0.4 \frac{GM}{R_h}.
\end{equation}
The exponential factor in eq.~[1] allows for the fact that the
protogalactic clouds do not form and virialize instantaneously, but require a
collapse time, $t_{\em vir} \sim 10^8$ yr.
We also assume that the mass of the protogalaxy is $\sim 4 \times 10^{11}
M_{\odot}$, and the initial mass of an average protogalactic fragment cloud is
$10^6 M_{\odot}$, following MS. If
the bulk of the mass of our Galaxy is in the form of non-baryonic dark matter,
one must allow for the fact that only a fraction of the $4 \times 10^{11}
M_{\odot}$ total mass is baryonic, but the essence of the model
remains. We will concentrate here on the
case of a purely baryonic halo, and also consider the possibility of
non-baryonic dark matter later in section 3.1.

The total stellar birthrate function is written as
\begin{equation}
\psi(t) = \alpha \psi_m + \beta \psi_c
\end{equation}
where $\psi_m$ represents the star birthrate induced by mergers, and is given
as
\begin{equation}
\psi_m \propto \lambda_m,
\end{equation}
and $\psi_c$ represents the quiescent star formation rate due to
self-regulated star formation and/or fragmentations of larger clouds, which is
taken to be
\begin{equation}
\psi_c \sim \rho_g^{1/2}.
\end{equation}
The age of the Galaxy is taken to be $\sim 15$ Gyr, and the collapse timescale
is
$\sim 6.1$ Gyr (MS).

There are several factors which enter into the formation of
globular clusters (Murray \& Lin 1993; Murray et al. 1993).
Following Murray et al. (1993),
we picture the protogalaxy as comprised of cool dense clouds in
pressure equilibrium with a hot intercloud medium from which they have
cooled.  The first factor which influences the clouds is that
they are subject to a Kelvin-Helmholtz (KH) instability which grows as the
clouds move through the background gas.  This instability leads
to the disruption of the clouds on a timescale
less than the dynamical collapse time unless their mass exceeds a critical
value.   Above this critical value,
the clouds are gravitationally stabilized and the timescale for the KH
instability becomes comparable to the dynamical time.
For clouds which survive the KH instability, two possibilities can occur.
Clouds with mass in excess of the Jeans mass are subject to a cooling
instability (Fall \& Rees 1985; Murray \& Lin 1989)
which leads to rapid star formation and fragmentation or
self-regulated star formation.  Neither scenario, however, will
lead to bound stellar clusters (Murray \& Lin 1992).
Globular clusters can form, however,
from clouds with masses above the critical value
(to survive the KH instability) but below the Jeans mass (to avoid
spontaneous fragmentation).  For our purposes this will correspond
to protocluster clouds (PCC's) with masses from $10^4$ to $10^6 M_{\odot}$
(Murray et al. 1993).

For clouds in this mass range three things can occur.
The clouds may be compressed, e.g. by cloud-cloud collisions.  If the
collision is strong enough ($v \ga 10$ km s$^{-1}$), then the shock
will trigger a thermal instability as it passes through a PCC
(Murray \& Lin 1989; Lin \& Murray 1992) and lead to the production
of a bound globular cluster.  Clouds which experience a collision
which is too weak, however, will coagulate and not collapse.
Clouds which do not collide are ultimately disrupted by the KH
instability.  Clearly, the
process of globular cluster formation is most efficient for clouds
with masses just below the Jeans mass ($\sim 10^6 M_{\odot}$)
which are the easiest to trigger into collapse, but which have
the longest lifetime against disruption by interaction with the
tenuous background medium.

Following MS, we envision the protogalaxy as initially comprised of
protogalactic clouds with mass near the Jeans mass with no subsequent
production of clouds in this mass range.  For such clouds, the
merger rate per cloud is given by eq.~[1], while the KH disruption
rate per cloud (Murray et al. 1993) is given by
\begin{equation} \label{lam,8}
\lambda_{\rm KH} \approx \frac{v}{R_{cl} D^{1/2}},
\end{equation}
where $R_{cl}$ is the cloud radius (taken here to be an average
tidal radius) and $D$ is the density ratio of the clouds relative to
the background medium (typically, $D \gg 1$).  Assuming pressure equilibrium
at the interface between the clouds and the background medium, the
density ratio is just related to the ratio of sound speeds
in the two media,
\begin{equation} \label{D,9}
D^{1/2} \approx \frac{c_1}{c_2} \approx \frac{T_1}{T_2}.
\end{equation}
The subscripts, 1 and 2,  refer to the background and clouds,
respectively.  We will assume that the cloud temperatures are radiatively
cooled to $10^4$ K.  The background temperature
will be just the dynamical temperature of the protogalaxy,
\begin{equation} \label{T,10}
T_1 = \frac{m_p v^2}{3 k}
\end{equation}
where $m_p$ is the proton mass.  For a typical virial
velocity of 200 km s$^{-1}$, $T_1 \sim 10^6$ K.

Thus, the KH disruption rate can be written as
\begin{equation}
\lambda_{\rm KH} = \frac{v}{R_{cl}} \frac{3kT_2}{m_p v^2} \approx 1.93
\times 10^5 \left(\frac{M_{tot}}{M_{\odot}}\right)^{-1/6}
\left(\frac{m}{M_{\odot}}\right)^{-1/3} \left(\frac{R_h}{\rm
kpc}\right)^{-1/2} \left(1-e^{-\frac{t-t_0}{t_{vir}}}\right)
{\rm Gyr}^{-1},
\end{equation}
taking into account the finite timescale for virializationOA. Here
$R_h$ denotes
the radius of the Galactic halo given by MS. The radius of the halo as a
function of time is plotted in Fig.\ 1.

In order to describe appropriately the evolution of the number and mass of
protogalactic clouds
(PGC's) represented by $N(t)$ and $m(t)$, the number of protocluster clouds
(PCC's) $N_{pcc}$,
the mass of which is $\sim 10^6 M_{\odot}$, and globular clusters
$N_{gc}$, we need to define the following quantities. First, we define an
auxiliary variable $f(t)$ as
\begin{equation}
f(t) \equiv \frac{m_0}{m(t)}.
\end{equation}
Subsequently we define
\begin{eqnarray}
L_1 (t) \equiv & \lambda_m \left\{1+\epsilon f_{\em coll}
f^{2/3}\frac{N_{pcc} (N_{pcc}-1)}{N(N-1)}\right\} & N_{pcc} > 1
\nonumber \\
 & \lambda_m & N_{pcc} \le 1,
\end{eqnarray}
\begin{eqnarray}
L_2 (t) \equiv &
\lambda_m\left(\frac{N}{N-1}\left(\frac{1+f^{1/3}}{2}\right)^2 +
\frac{f^{2/3}}{N-1} (N_{pcc}-2)\right\} & N_{pcc} > 2 \nonumber \\
 & \lambda_m \frac{N}{N-1} \left(\frac{1+f^{1/3}}{2}\right)^2  & N_{pcc}
\le 2,
\end{eqnarray}
and
\begin{eqnarray}
L_3 (t) \equiv & \lambda_m \left\{1-\epsilon f_{\em coll} f^{2/3}
\frac{N_{pcc} (N_{pcc}-1)}{N(N-1)}\right\} & N_{pcc} > 1 \nonumber
\\
 & \lambda_m  & N_{pcc} \le 1.
\end{eqnarray}
The above three quantities are basically merger rates modified by phase space
factors and the production of globular clusters, as we shall see later.
Here, $f_{\em coll}$ is the fraction of a Maxwell-Boltzmann cloud velocity
distribution with a relative velocity in excess of the trigger velocity
($\sim 10$ km s$^{-1}$).
Since the characteristic virial velocities (eq.~[2])
considered here are generally large compared to the trigger velocity, this
fraction can be taken as near unity. The factor $\epsilon$ is the efficiency
of globular cluster production.

With these definitions, we have first
\begin{equation}
\frac{dN}{dt} = - (L_1 + \lambda_{\rm KH})N,
\end{equation}
\begin{equation}
\frac{dN_{pcc}}{dt} = - (L_2 + \lambda_{\rm KH}^0)N_{pcc},
\end{equation}
where $\lambda_{\rm KH}^0$ is KH disruption rate with $m=10^6 M_{\odot}$, and
\begin{equation}
\frac{dN_{gc}}{dt} = \epsilon f_{\em coll} L_2 N_{pcc}.
\end{equation}

The interpretations are as follows: $L_1$
($L_2$) describes the decrease in number of PGC's (PCC's) due to mergers and
production of globular clusters. Since the PGC system consists of clouds with
a spectrum of masses with the average mass of $m(t)$ while the PCC's are clouds
of
mass $m_0 = 10^6 M_{\odot}$ by fiat, the merger rates and KH disruption rates
for the evolution of the two systems are different. The globular cluster
production rate is described by the efficiency ($\epsilon$), the
Maxwell-Boltzmann
factor ($f_{\em coll}$), and the merger rate ($L_2 N_{pcc}$). Since we
do not know the efficiency for globular cluster production via
this mechanism, we leave it as a free parameter to be determined by other
constraints.

With these equations, and the supposition that no new PCC's are formed after
their initial appearance, the number of PGC's and PCC's at time $t$ are just
\begin{equation}
N(t) = N(0) \exp\left(-\int^{t}_{t_0} (L_1 + \lambda_{\rm KH}) dt'\right),
\end{equation}
and
\begin{equation}
N_{pcc} (t) = N(0) \exp\left(-\int^{t}_{t_0} (L_2 + \lambda_{\rm KH}^0)
dt'\right).
\end{equation}

The mass evolution equations are modified from those in MS as
\begin{equation}
\frac{dm_g}{dt} = \left(L_1- (1-R)\psi - \epsilon f_{\em coll} L_2
\frac{N_{pcc}}{N}\right) m_g,
\end{equation}
and
\begin{equation}
\frac{dm_{\ast}}{dt} = -\left(L_3 + \epsilon f_{\em coll} L_2
\frac{N_{pcc}}{N}\right) m_{\ast} + (1-R)\psi m_g.
\end{equation}
The first term in the braket in eq.~[19] describes the mass increase due to
mergers, the second term mass loss by star formation, and the third term
mass loss by production of globular clusters.
Eq.~[20] can be interpreted in a similar way.

Therefore, eqs.~[14], [15], [16], [19], and [20] are the evolution equations
that describe the Galactic halo system.

\section{Results}
\subsection{Main results}

There are two different points of view as to the mechanism by which globular
clusters obtain their metallicities: self-enrichment schemes and previously
enriched environments.
The fact that the
metallicity within most globular clusters is very uniform and that the
metallicity of clusters
is
not much different from that of field stars (see Zinn 1988) argues in favor of
environment as fixing the metallicity.
Therefore, we adopt the latter view, so we identify the metallicity
of the field as that of globular clusters, although
self-enrichment may be possible (e.g. Brown et al. 1991).
We reproduce the age-metallicity relation of globular clusters and show it in
Fig\ 2. It is compared with the data of Twarog (1980) (see also Colin,
Schramm, \& Peimbert 1994).

Figs.\ 3 shows the rate of globular cluster production as a
function of time, and Fig.~4 shows the metallicity distribution. For these
figures, $\epsilon = 10^{-3}$.
In Fig.~4 $dN_{gc}/dZ$ is compared directly to Zinn's data (1985).
{}From Fig.\  3, we observe a very distinct
epoch of globular cluster formation at the early stage of
galaxy formation. Although there exists a second peak at the time
of the collapse of the protogalaxy, it is smaller than the first peak by about
a factor of $10^7$ (we speculate in section 3.2
about processes that might enhance this secondary peak).
As for Fig.\ 4, one can see clearly that the predicted curve is almost
identical to the observed metallicity
distribution except for the fact that there is no pronounced second peak that
corresponds to disk clusters (Zinn 1985, 1988).
We will return to the second peak later.

The shape of these curves is unique, i.e. they are insensitive to values
of the only free parameter in our model $\epsilon$. Although $\epsilon$ enters
the
equations in a nontrivial way, it does not affect the overall behavior of the
system as long as it remains a small parameter. The distribution is largely
fixed by the initial burst of globular cluser formation due to the high merger
rate when the density of PCC's is highest just after formation. The formation
rate then decreases as the protogalactic halo expands (MS). Thus, $\epsilon$
changes only
the overall amplitude of the globular cluster production, and it can be more
or less fixed by the present number of globular clusters. Fig.\ 5 shows
the metallicity distribution curves as $\epsilon$ changes from 10$^{-1}$ to
10$^{-3}$.
Table 1 lists values for the final numbers of
globular clusters $N_{gc} (T)$, and the coefficients for the quiescent
star formation $\beta$ adjusted to give the present local gas mass fraction of
$\mu_g \simeq
0.28$ for different choices of $\epsilon$.

Since the
number of globular clusters in our Galaxy is observed to be $\sim 100$ to
150, then
$\epsilon \sim 10^{-3} $ if no significant GC destruction has occurred over the
Galactic lifetime.

In addition, our results argue that halo clusters should have a relatively
small age spread of less than 1 Gyr and a relatively large metallicity spread.
Sandage (1993) determined the ages and the metallicities of 24 Galactic
globular
clusters. All but one are halo clusters according to Zinn's classification
scheme. He reported that the ages of those clusters are $14 \pm 1.5$ Gyr
on the average, and also noted that these clusters have quite small age spread
compared to the metallicity spread, which implies that the (halo) globular
clusters formed in a short timespan at the very early stage of galaxy
formation and the metallicity buildup of our Galaxy was quite fast. This may
be regarded tentatively as supporting the conclusion of this model (see also
Lee, Demarque, \& Zinn 1985).

We point out, however, that measurements of globular age and metallicity have
large uncertainties (see Sandage 1993; Chaboyer et
al. 1992a, 1992b; VandenBerg, Bolte, \& Stetson 1990).

Figs.\ 6 and 7 shows the numbers of clouds ($N, N_{pcc}, N_{gc}$) and
masses ($m(t), m_g, m_{\ast}$) as functions of time.

We adopted 15 Gyr for the age of our Galaxy as quoted in MS, but since there
is a possibility that this might be reduced by as much as a few Gyr (Shi,
Schramm, \& Dearborn
1994), we examined a few
different ages for the Galaxy. As one can find in Fig.\ 8, the metallcity
distribution curve remains virtually unchanged.

We can also consider the case of possible nonbaryonic dark matter (NBDM). If we
include NBDM with fixed total mass, we have less baryons than the case of the
purely baryonic halo. So we increase the efficiency ($\epsilon$) to
$10^{-2}$ compared with 10$^{-3}$ for purely baryonic halo case. Fig.\ 9 shows
the
metallcity distribution curve with 90 \% non-baryonic dark matter in the halo.
We
note that the maximum shifts towards slightly higher metallicity. The purely
baryonic halo case, is thus slightly more desirable than NBDM model.

\subsection{Absence of the second peak}

As was mentioned above, there was no appreciable second peak in the metallicity
distribution curve of globular clusters that correspond to disk clusters with
higher metallicity. The reason that we didn't get the second peak within the
framework of our model is rather straightforward. PCC's, which are progenitors
of globular clusters, disappear quite rapidly within a few Gyr of galaxy
formation, as one can see in Fig.\ 6. They turn into globular clusters by
starbursts triggered by collisions with other PGC's, merge and coagulate with
other clouds, or get disrupted due to KH instability moving through the hot
medium. Since the decrease of the number of PCC's is quite rapid, there are not
many PCC's when the protogalaxy finally collapses to form the disk. Thus, even
though we have a high merger rate again at the time of collapse there is no
significant production of globular clusters by this mechanism unless some way
is envisaged to prevent total PCC destruction or to form new PCC's during
collapse.

However, it is still suggestive that the mean metallicity of the observed
second peak, i.e. disk clusters ([Fe/H] $\simeq -0.45$) roughly coincides
with the end of the collapse epoch ($t \sim 6$ Gyr) in our model.
Therefore, we argue that the disk clusters might form more efficiently than
considered here.
It seems quite likely, however, that the mechanism that
causes the production of disk clusters is associated with the
collapse of the protogalaxy. This would require
subsequent production of PCC's during or even after the collapse of the
protogalaxy. So this indicates that the two
globular cluster populations might have different origins, although they are
both triggered by violent motions of gas clouds which induce
mergers and collapse.

\subsection{Globular clusters in other galaxies: specific frequency}

It might be also worthwhile to consider globular clusters in other galaxies.
As Ashman \& Zepf (1992) already pointed out, elliptical galaxies tend to
have many more globular clusters than the same size spirals. If we accept the
possibility of mergers of spiral galaxies as one of the leading causes for the
formation of elliptical galaxies, this presents a strong case for mergers as
sources
of globular cluster formation. Ashman \& Zepf (1992) also noted, however, that
the specific
frequency (the number of globular clusters per galaxy luminosity) for a given
morphology is almost independent of the mass of the host galaxy.
{}From our calculation we can estimate the approximate dependence of the
globular cluster production rate on galactic parameters.

Since the globular cluster production is dominant early on, we need to examine
the production rate very early on. From eq.~[16] and using the functional
approximations used in MS, we have

\begin{equation}
\frac{dN_{\em gc}}{dt} \propto R_h^{-3/2} M^{11/6}.
\end{equation}
If we suppose
\begin{eqnarray}
R_h \propto M^{\mu} & ( \mu > 0 ), \nonumber
\end{eqnarray}
then
\[
\frac{dN_{\em gc}}{dt} \sim M^{\nu}
\]
where
\begin{equation}
\nu \simeq -\frac{3}{2} \mu + \frac{11}{6}.
\end{equation}
The specific frequency of globular clusters for a given host galaxy with mass
$M$, is defined as the number of globular clusters per luminosity. If we
suppose that luminosity is proportional to the mass of a galaxy, we have
\begin{eqnarray}
S \equiv \frac{N_{\em gc}}{L} \propto \frac{N_{\em gc}}{M} \sim \frac{dN_{\em
gc}/dt}{M} \propto M^{\sigma}. &  ( \sigma = \nu -1 ) \nonumber
\end{eqnarray}
where
\begin{equation}
\sigma \simeq -\frac{3}{2}\mu + \frac{5}{6}.
\end{equation}
If radius scales with mass, i.e. $M \propto R^3$, or equivalently $\mu =
1/3$, we have
\begin{equation}
\sigma \simeq \frac{1}{3}.
\end{equation}
{}From the above relation, we conclude that the specific frequency $S$ depends
weakly, if at all, on the mass of the galaxy. Fig.\ 10 shows the
numerical evaluation of the specific frequency as a function of the galaxy
mass for a reasonable range. Actual numerical results confirmed that $S$
indeed depends very weakly on the mass. We find $\sigma \sim 0.12$, which is
somewhat smaller than the above analytic estimate.
We would like to caution, however, that this is
an extremely crude estimate because we assume the same evolution history for
galaxies and we do not take into account other mechanisms that affect
globular clusters (i.e. subsequent production of disk clusters, disruptions,
tidal captures, etc.; Ostriker 1988; Spitzer 1987).

\section{Discussion}

{}From the merger model of MS, a model for globular cluster formation arises
quite naturally, and we can explain many characteristics of the halo
population of globular
clusters in our Galaxy in a consistent manner. In particular, we successfully
predict the metallicity distribution for globular clusters.
This model has the added attraction that it will be
testable in the near future and the predictions of the model are robust, i.e.
insensitive to choices of parameters. Thus, the merger model seems to be a
viable and
self-consistent model for galaxy formation.

Previously in MS, we used the instantaneous recycling approximation to
simplify the analysis.
However, if we abandon the instantaneous recycling approximation, we should
employ an appropriate initial mass function (IMF), and the calculations will
be improved somewhat
(Mathews \& Schramm 1994). But
the essential features of the model and the prediction in the paper will not
change appreciably.

Recently there were observations by ROSAT about the mass fraction of hot gas
in rich clusters like Coma cluster (Briel, Henry, \& B\"{o}hringer 1992;
Henriksen \& Mamon
1993
for example), and the observed ratio seems to be large
considering the scale of the clusters sampled (White et al. 1993) and the
constraints on baryonic density from Big Bang Nucleosynthesis (Copi,
Schramm, \& Turner 1994). The merger
model in the context of clusters of galaxies rather than single galaxies might
naturally give a
plausible explanation for the ROSAT observations as well, and this is
investigated in detail in a following paper (Mathews, Charlot, \& Schramm
1994).
\vskip 0.5in

\noindent{We} would like to thank Sydney van den Bergh for some interesting
questions which helped prompt the present paper. We also thank James Truran,
Stephen Murray, Douglas Lin, Xiangdong Shi, Geza Gyuk, Bruce
Carney, and Pedro Colin for very helpful
conversation. The work is performed in part under the auspices of the U.S.
Department of Energy by Lawrence Livermore National Laboratory under Contract
number W-7405-ENG-48 and Nuclear Theory Grant Number SF-ENG-48.
The work is supported at the University of Chicago in part by the DOE, by NASA
and by the NSF, and at Fermilab by the DOE and through NASA Grant NAGW 2381.
SL acknowledges the support of POSCO Scholarship Foundation in Korea.
\newpage
\begin{table}[h]
\begin{center}
\begin{tabular}{ccc} \hline
$\epsilon$ & $N_{gc}$ & $\beta$ \\ \hline\hline
$10^{-3}$ & $3.135 \times 10^2$ & $1.985 \times 10^{-4}$ \\
$10^{-2}$ & $3.134 \times 10^3$ & $1.969 \times 10^{-4}$ \\
$10^{-1}$ & $3.127 \times 10^4$ & $1.792 \times 10^{-4}$ \\
1 & $3.020 \times 10^5$ & $1.478 \times 10^{-4}$ \\ \hline
\end{tabular}
\end{center}
\caption{Calculated present number of globular clusters $N_{gc}$
and the coefficient of quiescent start formation rate $\beta$ in units of
Gyr$^{-1} (M_{\odot}/{\rm kpc}^3)^{-1/2}$.}
\end{table}

\newpage
\newpage

\section*{References}
\frenchspacing
\noindent{Ashman, K. M., \& Zepf, S. E. 1992, ApJ, 384, 50.}\\
\noindent{Briel, U. G., Henry, J. P., \& B\"{o}hringer, H. 1992, A\&A, 259,
L31.}\\
\noindent{Brown, J. H., Burkert, A., \& Truran, J. W. 1991, ApJ, 376,
115.}\\
\noindent{Chaboyer, B., Deliyannis, C. P., Demarque, P., Pinsonneault, M. H.,
\& Sarajedini, A. 1992a, ApJ, 388, 372.}\\
\noindent{Chaboyer, B., Sarajedini, A., \& Demarque, P. 1992b, ApJ, 394,
515.}\\
\noindent{Colin, P., Schramm, D. N., \& Peimbert, M. 1994, ApJ, 426, 459.}\\
\noindent{Copi, C., Schramm, D. N., \& Turner, M. S. 1994, Science,
submitted.}\\
\noindent{Eggen, O. J., Lynden-Bell, D., \& Sandage, A. R. 1962, ApJ, 136,
748.}\\
\noindent{Fall, S. M., \& Rees, M. J. 1985, ApJ, 298, 18.}\\
\noindent{Henriksen, M. J., \& Mamon, G. A. 1994, ApJL, in press.}\\
\noindent{Lin, D. N. C., \& Murray, S. D. 1992, ApJ, 394, 523.}\\
\noindent{Mathews, G. J., Charlot, S., \& Schramm, D. N. 1994, in
preparation.}\\
\noindent{Mathews, G. J., \& Schramm, D. N. 1993, ApJ, 404, 468 (MS).}\\
\noindent{Mathews, G. J., \& Schramm, D. N. 1994, in preparation.}\\
\noindent{Murray, S. D., \& Lin, D. N. C. 1989, ApJ, 339, 933.}\\
\noindent{Murray, S. D., \& Lin, D. N. C. 1992, ApJ, 400, 265.}\\
\noindent{Murray, S. D., \& Lin, D. N. C. 1993, in Globular Cluster-Galaxy
Connection, eds. G.~H.~Smith, \& J.~P.~Brodie (San Francisco: Astronomical
Society of the Pacific), 738.}\\
\noindent{Murray, S. D., White, S. D. M., Blondin, J. M., \& Lin, D. N. C.
1993,
ApJ, 407, 588.}\\
\noindent{Ostriker, J. 1988 in The Harlow-Shapley Symposium on Globular
Cluster Systems in Galaxies, eds. J.~E.~Grindlay, \&  A.~G.~D.~Philip
(Dordrecht: Kluwer), 271.}\\
\noindent{Peebles, P. J. E., \& Dicke, R. H. 1968, ApJ, 154, 891.}\\
\noindent{Sandage, A. 1993, AJ, 106, 719.}\\
\noindent{Searle, L., \& Zinn, R. 1978, ApJ, 225, 357.}\\
\noindent{Shi, X., Schramm, D. N., \& Dearborn, D. 1994, Physical Review D,
50, in press (August 15).}\\
\noindent{Spitzer, L. 1987, Dynamical Evolution of Globular Clusters
(Princeton: Princeton Univ. Press), 117.}\\
\noindent{Twarog, B. A. 1980, ApJ, 242, 242.}\\
\noindent{VandenBerg, D. A., Bolte, M., \& Stetson, P. B. 1990, AJ, 100,
445.}\\
\noindent{White, S. D. M., Navarro, J. F., Evrard, A. E., \& Frenk, C. S.
1993, Nature,  366, 429.}\\
\noindent{Zinn, R. 1985, ApJ, 293, 424.}\\
\noindent{Zinn, R. 1988 in The Harlow-Shapley Symposium on Globular
Cluster System in Galaxies, eds. J.~E.~Grindlay, \&  A.~G.~D.~Philip
(Dordrecht: Kluwer), 37.}\\
\noindent{Zinn, R. 1993, in Globular Cluster-Galaxy Connection, eds.
G.~H.~Smith, \& J.~P.~Brodie (San Francisco: Astronomical Society of the
Pacific), 38.}\\

\newpage
\section*{Figure Captions}
\noindent{\bf Figure 1.}\\
\noindent{Radius of the protogalaxy $R_h (t)$ as a function of time (MS).}\\
\noindent{\bf Figure 2.}\\
\noindent{Metallicity Z as a function of time. Data points are from
Twarog (1980).}\\
\noindent{\bf Figure 3.}\\
\noindent{The formation rate of globular clusters $dN_{gc}/dt$ as a function
of time with $\epsilon = 10^{-3}$.}\\
\noindent{\bf Figure 4.}\\
\noindent{Calculated formation rate of globular clusters $dN_{gc}/dZ$ as a
function
of metallicity (line) with $\epsilon = 10^{-3}$, compared with data
(histogram) from Zinn
(1985).}\\
\noindent{\bf Figure 5.}\\
\noindent{The formation of globular clusters $dN_{gc}/dZ$ as a function
of metallicity for different choices of $\epsilon$. Values of $\epsilon$ are
$10^{-3}, 10^{-2}, 10^{-1},$ and $1$ from left to right. $\epsilon=10^{-2}$
case is
hardly distinguishable from $\epsilon=10^{-3}$ in the figure.}\\
\noindent{\bf Figure 6.}\\
\noindent{Calculated number of protogalactic clouds $N$ (solid curve),
protocluster
clouds $N_{\em pcc}$ (dotted), and globular clusters $N_{gc}$ (dashed) as
functions of time.}\\
\noindent{\bf Figure 7.}\\
\noindent{The average mass $m$ of a PGC (top curve), the average mass $m_g$ in
gas of a PGC (middle curve), and the
average mass $m_{\ast}$ in stellar remnant of a PGC (bottom curve) as
functions of time.}\\
\noindent{\bf Figure 8.}\\
\noindent{$dN_{\em gc}/dZ$ as a function of Z for different values of the
age of the Galaxy. The solid curve corresponds to 15 Gyr for the age of the
Galaxy, the dotted curve 13 Gyr, and the dashed curve 11 Gyr respectively.}\\
\noindent{\bf Figure 9.}\\
\noindent{$dN_{gc}/dZ$ as a function of Z, with and without non-baryonic dark
matter in the halo. $\epsilon = 10^{-3}$ for the purely baryonic case (solid
curve), and
$\epsilon = 10^{-2}$ for the non-baryonic case (dotted curve).}\\
\noindent{\bf Figure 10.}\\
\noindent{The specific frequency $S$ of globular clusters as a function of the
galaxy mass,
assuming  $\mu = 1/3$ and $\epsilon = 10^{-3}.$}\\

\end{document}